\documentclass[pss,fleqn]{w-art}
\usepackage{times}
\usepackage{w-thm}
\usepackage[]{epsfig,graphicx}
\begin{document}
\DOIsuffix{theDOIsuffix}
\Volume{XX}
\Issue{1}
\Month{01}
\Year{2006}
\pagespan{3}{}
\Receiveddate{9 September 2006}
\keywords{superconductivity}
\subjclass[pacs]{ 74.20.Rp, 74.20.-z, 74.25.Bt
}




\title[Magnetic field 
effect on  a p-wave superconductor]{
Magnetic field orientation
effect on specific heat in a p-wave superconductor
}


\author[G. Litak]{Grzegorz Litak\footnote{Corresponding
     author: e-mail: {\sf g.litak@pollub.pl}, Phone:
+48\,81\,5381573,
     Fax: +48\,81\,5250808} \inst{1}}
\address[\inst{1}]{Department of Mechanics, Technical University of
Lublin,
Nadbystrzycka 36,
PL-20-618 Lublin, Poland}
\author[K.I. Wysoki\'nski]{ Karol I. Wysoki\'nski \inst{2}}
\address[\inst{2}]{Institute of Physics,
M. Curie-Sk\l{}odowska University,
ul. Radziszewskiego 10, 20-031 Lublin, Poland}
\begin{abstract}
The effect of magnetic field on specific heat of
a p-wave spin triplet superconductor has been analyzed. 
To describe superconductivity we used a single band model
with various realizations of p-wave order parameter
and analyzed the effect of a Doppler shift in different temperatures.
\end{abstract}
\maketitle                   

\section{Introduction} 
The discovery of superconductivity is strontium ruthenate \cite{maeno1994} 
and subsequent proposal of its spin triplet odd parity symmetry \cite{maeno2001}
has triggered a lot of experimental and theoretical activities \cite{mackenzie2003}.
One of the issues of great importance is the identification of the pairing
symmetry of this superconductor. The expected sizable spin-orbit coupling
and the temperature independent Knight shift measured with magnetic field 
in the basal a-b plane \cite{ishida1998} suggest the realization of the state 

\begin{figure}[ht]
 \epsfysize=4cm
 \centerline{\epsfig{file=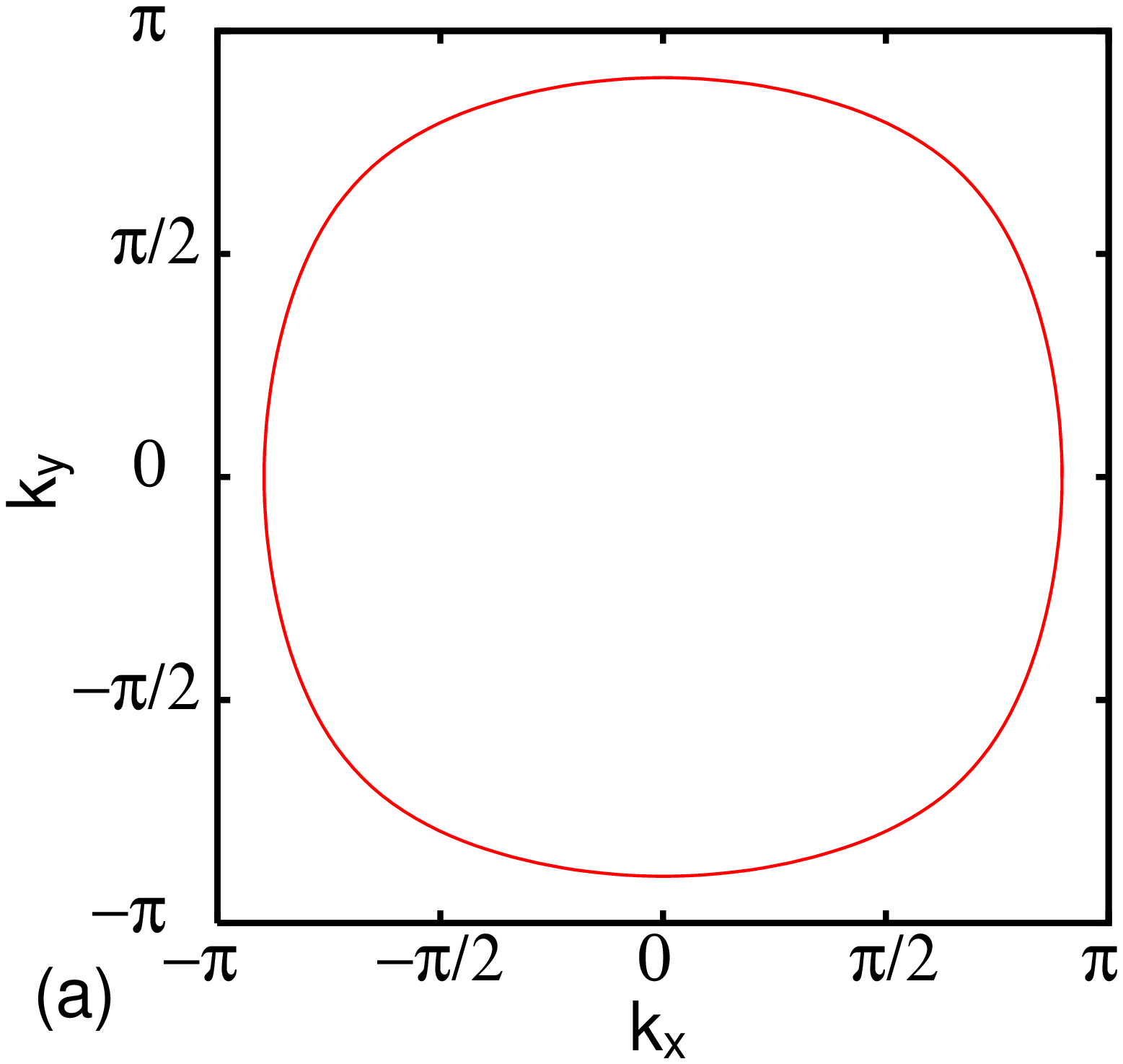,width=6.0cm,angle=-90}
\epsfig{file=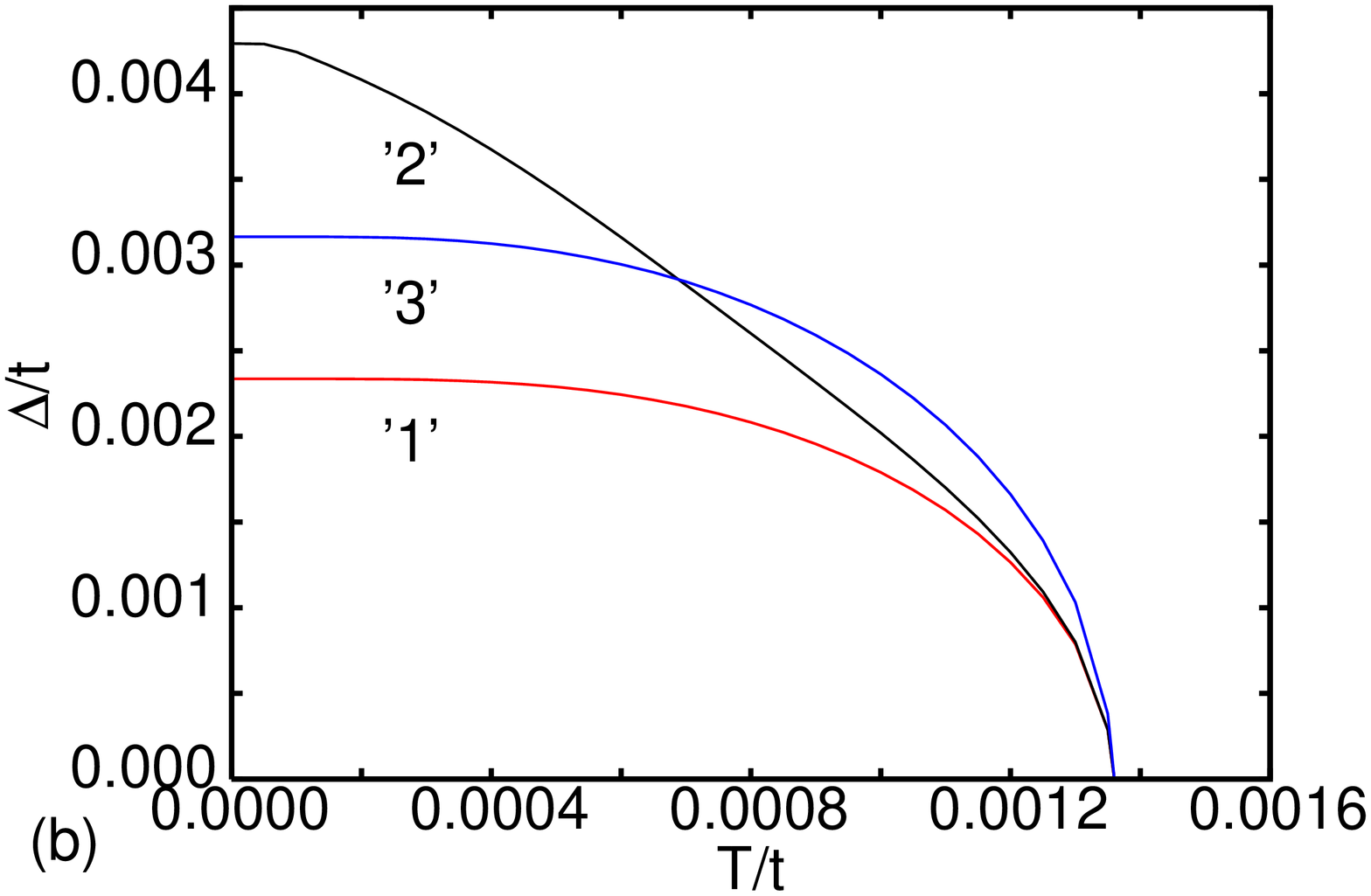,width=6.0cm,angle=-90}
}
\vspace{-0.2cm}
 \caption{
(a) Fermi surface of the gamma band Sr$_2$RuO$_4$. (b)
Pairing potential for three different p-wave solutions $\Delta ({\bf k}) = \Delta_x 
\sin k_x + \Delta_y \sin k_y$:
'1' complex ($\Delta_y= i \Delta_x$), '2' real
($\Delta_y= \Delta_x$), and '3' dipole  ($\Delta_x \neq 0$, $\Delta_y = 0$). System parameters in the electron 
nearest 
neighbour hopping unit $t$:
next nearest neighbour hopping $t'/t=0.45$, nearest neighbour attraction $U/t$=-0.446, chemical 
potential $\mu=1.5966$.
 \label{fig1}}
\vspace{-0.2cm}
\end{figure}

\begin{figure}[ht]
 \epsfysize=4cm
\centerline{ \hspace{1cm} \epsfig{file=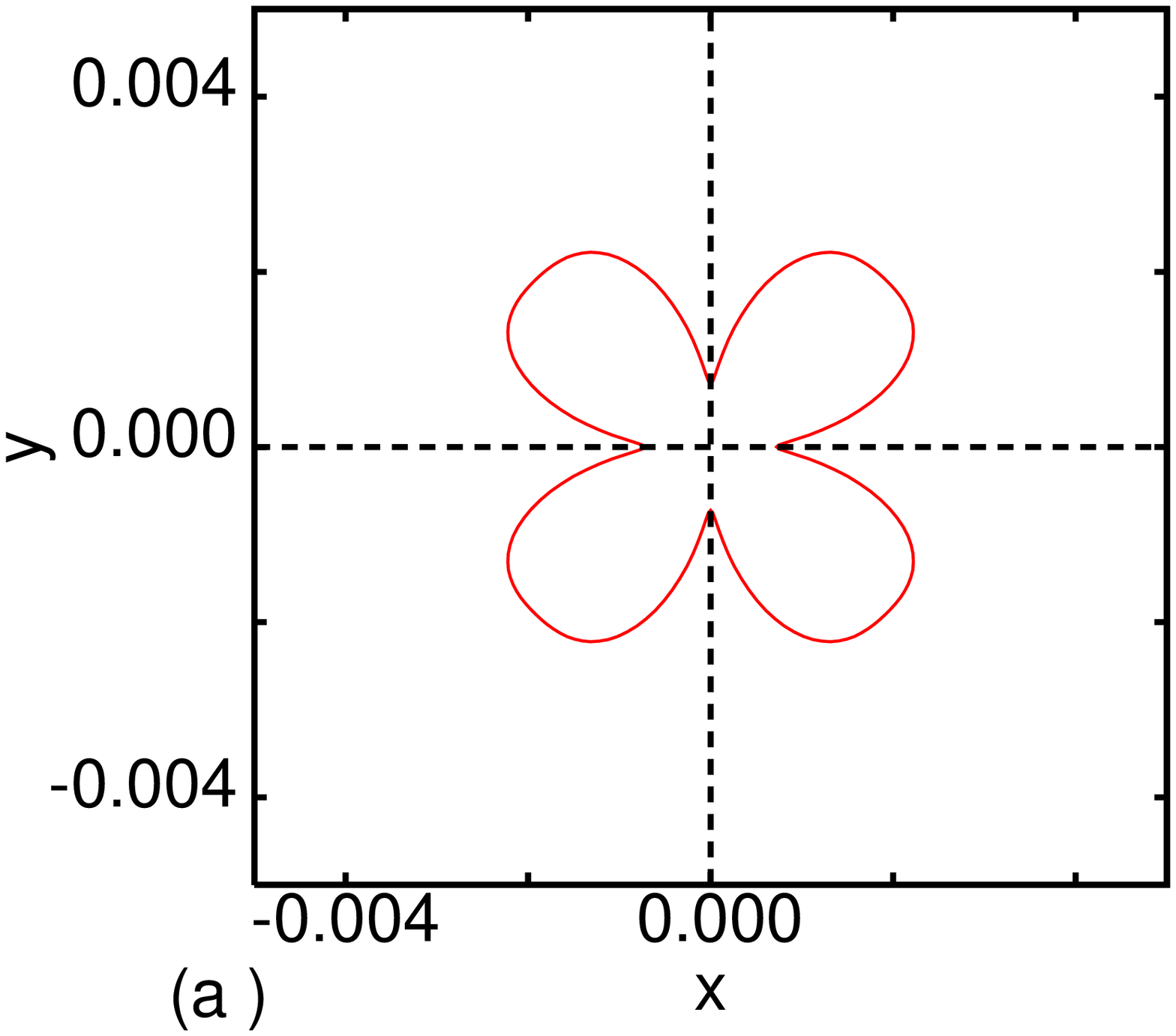,width=5.5cm,angle=-90}
\hspace{-2.3cm}
\epsfig{file=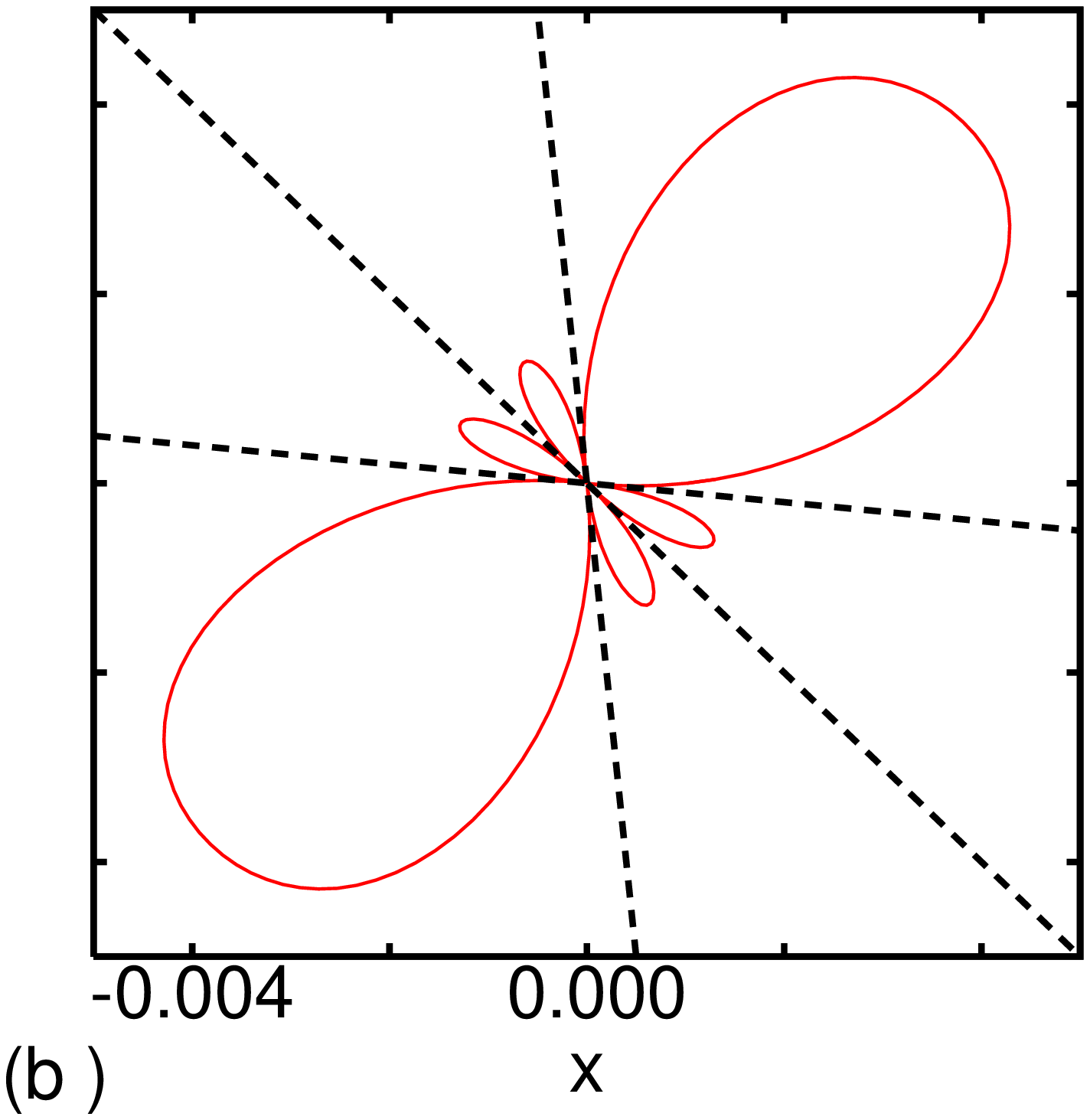,width=5.5cm,angle=-90} \hspace{-2.3cm}
\epsfig{file=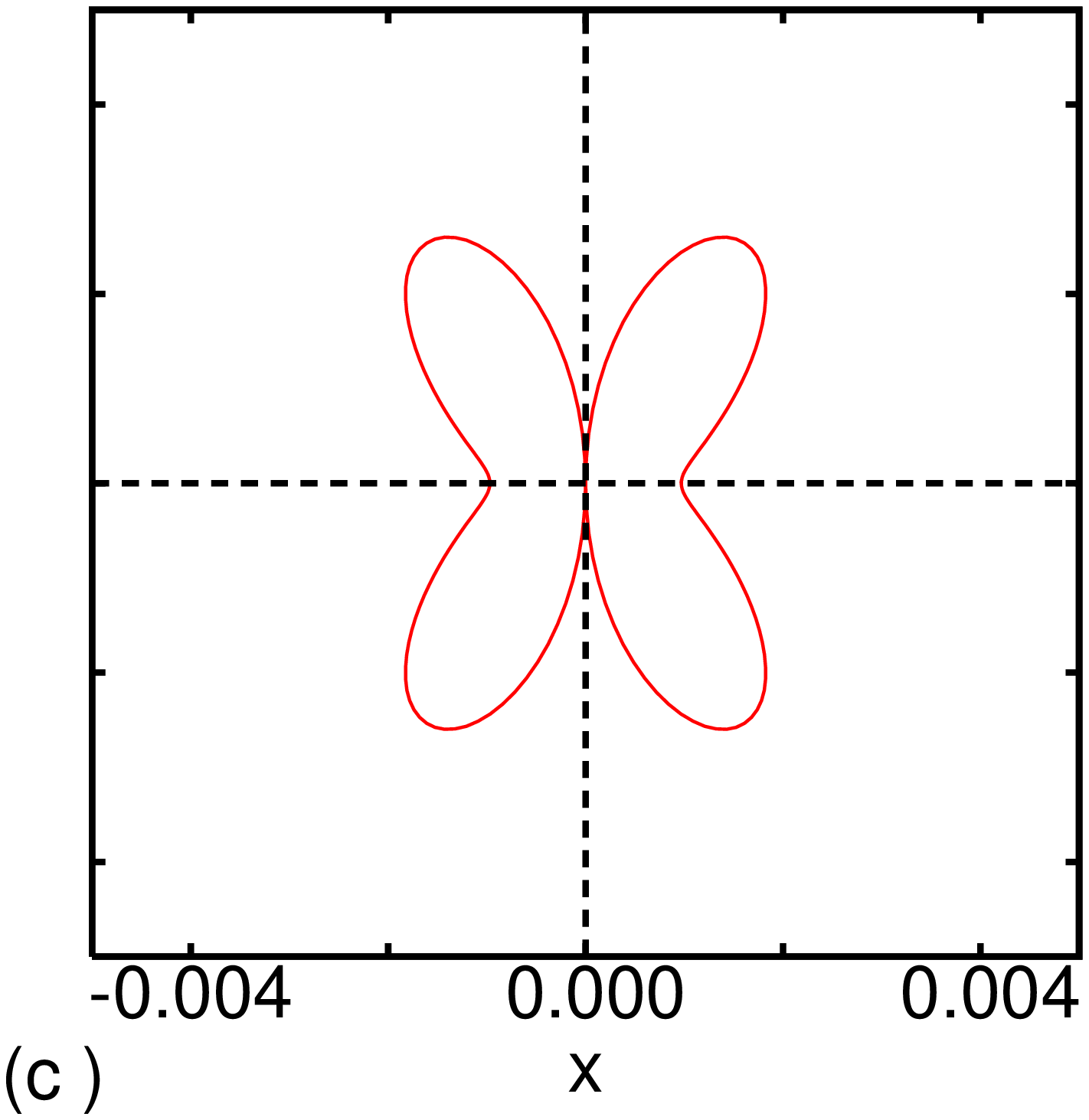,width=5.5cm,angle=-90}
}
\caption{
 Eigenvalues at the Fermi surface
$E_k(E_f)$ (measured from the Fermi surface) for complex, real and dipole solutions,  ((a), (b) and 
(c)
as in Fig.
\ref{fig1}b '1', '2' and '3', respectively)
calculated in the temperature $T/t=0.006$. Dashed lines
correspond to lines of zeros or minima in eigenvalues at the Fermi surface.
 \label{fig2}}
\vspace{-0.2cm}
\end{figure}

\begin{equation}
  {\bf d}({\bf k}) = \Delta (\sin{k_x}+i\sin{k_y})\hat{\bf e}_z,
  \label{eq1}
\end{equation}
which is chiral and has d-vector pointing in z-direction.

The chiral state (Eq. \ref{eq1}) has been deduced
\cite{deguchi2004a,deguchi2004b} from
specific heat measurements on Sr$_2$RuO$_4$ in the presence of the magnetic field. 
To shed some light on the issue
we shall study the magnetic field effect on the specific heat of 
a superconductor with different order parameters.
To describe the various experiments on strontium ruthenate 
one needs a 3-dimensional, 3-band model \cite{annett2002}.
In this work we shall study a fairly simpler
model consisting of a single band of a two-dimensional
character \cite{micnas1990,litak2000,litak2002} 
where the magnetic field is applied at an angle $\phi$ with respect
to the $x$-axis, 
 and study its effect on the specific heat.
The spectrum we start with corresponds to the $\gamma$ band in Sr$_2$RuO$_4$ 
\cite{annett2002}.
However, we neglect  the two other bands  and limit our consideration to 
two dimensions. Our approach of introducing the magnetic field is similar to
that of  Tanaka {\it et al.} \cite{tanaka2003}, who studied the 
influence of an in-plane magnetic field on the thermal conductivity
of the same material with the spectrum approximated 
by  the $\alpha$ and $\beta$ bands.

The study here allows a better
understanding of the influence of a magnetic field and of temperature on 
the specific heat of superconductors with different order parameters.

\section{ The model} 

\begin{figure}[ht]
 \epsfysize=4cm
 \centerline{\epsfig{file=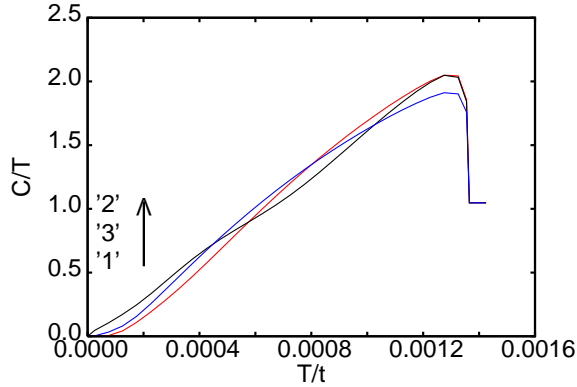,width=6.0cm,angle=-90}}
\vspace{-0.2cm}
 \caption{
   Specific heat $C$ over temperature $T$ versus temperature $T$ for
three different p-wave solutions
(denoted by '1', '2', and '3' as in Fig. 1b).
\label{fig3}}
\vspace{-0.2cm}
\end{figure}

The system is described by negative $U$ Hubbard model \cite{micnas1990} with 
the Hamiltonian

\begin{eqnarray}
 H = \sum_{ij\sigma} \left(t_{ij} - \mu \delta_{ij}\right)
     c^+_{i\sigma} c_{j\sigma} +
     \frac{1}{2} \sum_{ij\sigma,\sigma'} U^{\sigma\sigma'}_{ij} n_{i\sigma} n_{j\sigma'},
 \label{eq2}
\end{eqnarray}
where $i$, $j$ label sites of a square lattice, $t_{ij} = -t$ is the hopping
integral between nearest-neighbour sites, $\sigma$ is the electron spin 
and $\mu$ is the chemical potential.
$U^{\sigma\sigma'}_{ij} < 0$ describes attraction between electrons with spins 
$\sigma$ and $\sigma'$ occupying
 sites $i$ and $j$, respectively. We have assumed the hopping parameter  $t$ 
to be our energy unit. 
In Fig. (\ref{fig1}a) we show the Fermi circle of the $\gamma$ band of interest. 
 
The mean-field type decoupling of the interaction, together with Bogolubov-Valatin
quasi-particle transformation have been used to derive the superconducting
equations
\begin{equation}
 \sum_{j\sigma'} \left(\begin{array}{c}
 E^\nu - H(ij);  \hspace{0.5cm}
 \Delta^{\sigma\sigma'} (ij)\\
 \Delta^{* \sigma'\sigma} (ij);  \hspace{0.5cm}
 E^\nu +  H (ij)
\end{array}\right)
\left(\begin{array}{ll}
 u^\nu_{j\sigma'}\\
v^\nu_{j\sigma'}\end{array}\right)=0\,, \label{eq3}
\end{equation}
where  $ H (ij) $
is the normal spin independent part of the Hamiltonian, and
the $\Delta^{\sigma\sigma'}(ij)$ is
self consistently given
in terms of the pairing amplitude, or order parameter,
$\chi^{\sigma\sigma'}(ij)$,
\begin{equation}
 \Delta^{\sigma\sigma'} (ij) = U^{\sigma\sigma'}(ij)
\chi^{\sigma\sigma'}(ij)\,. \label{eq4}
\end{equation}
defined by the usual relation
\begin{equation}
\chi^{\sigma\sigma'}(ij) =
\sum_{\nu} u^\nu_{i\sigma}v^{\nu *}_{j\sigma'}
(1 - 2f(E^\nu))\,, \label{eq5}
\end{equation}
where $\nu$ enumerates the solutions of Eq.~\ref{eq3}.

We have solved the  Bogolubov-deGennes equations for the 
temperature-dependent superconducting
order parameter and the results for 3 different order parameters
are shown in Fig. (\ref{fig1}b).  Note the difference between curve '2' and curves 
'1', '3'. As is plotted here, 
the amplitudes  of these differences do not mimic the average value of pairing potential 
which
would not differer so much. For better clarity we show the angular dependence of 
eigenvalues in Fig. \ref{fig2}. 
Straight lines
correspond to nodal lines or minima of eigenvalues on the Fermi surface.
Note that only the complex solution has a 4-fold symmetry (Fig. \ref{fig2}a) , while 
the dipole and 
real solutions have a  2-fold symmetry (Fig. \ref{fig2}b-c). Moreover the real 
solution is the most anisotropic with respect to the angle and possesses three 
nodal lines. 
On the other hand the complex solution has no line of nodes and the dipole has one.

The low-temperature specific heat $C$ across 
the superconducting state
has been calculated using the formula \cite{litak2004}

\begin{equation}
C= - 2 k_B \beta^2 \frac{1}{N}\sum_{k} E_k \frac{ \partial f(E_k)}{\partial \beta},
\label{eq6}
\end{equation}
where $\beta$ denotes the inverted temperature $1/(k_BT)$ and $f(E_k)$ is the Fermi 
function defined for the eigenvalue $E_k$. 

\begin{figure}[ht]
 \epsfysize=4cm
 \centerline{ \hspace{1cm} \epsfig{file=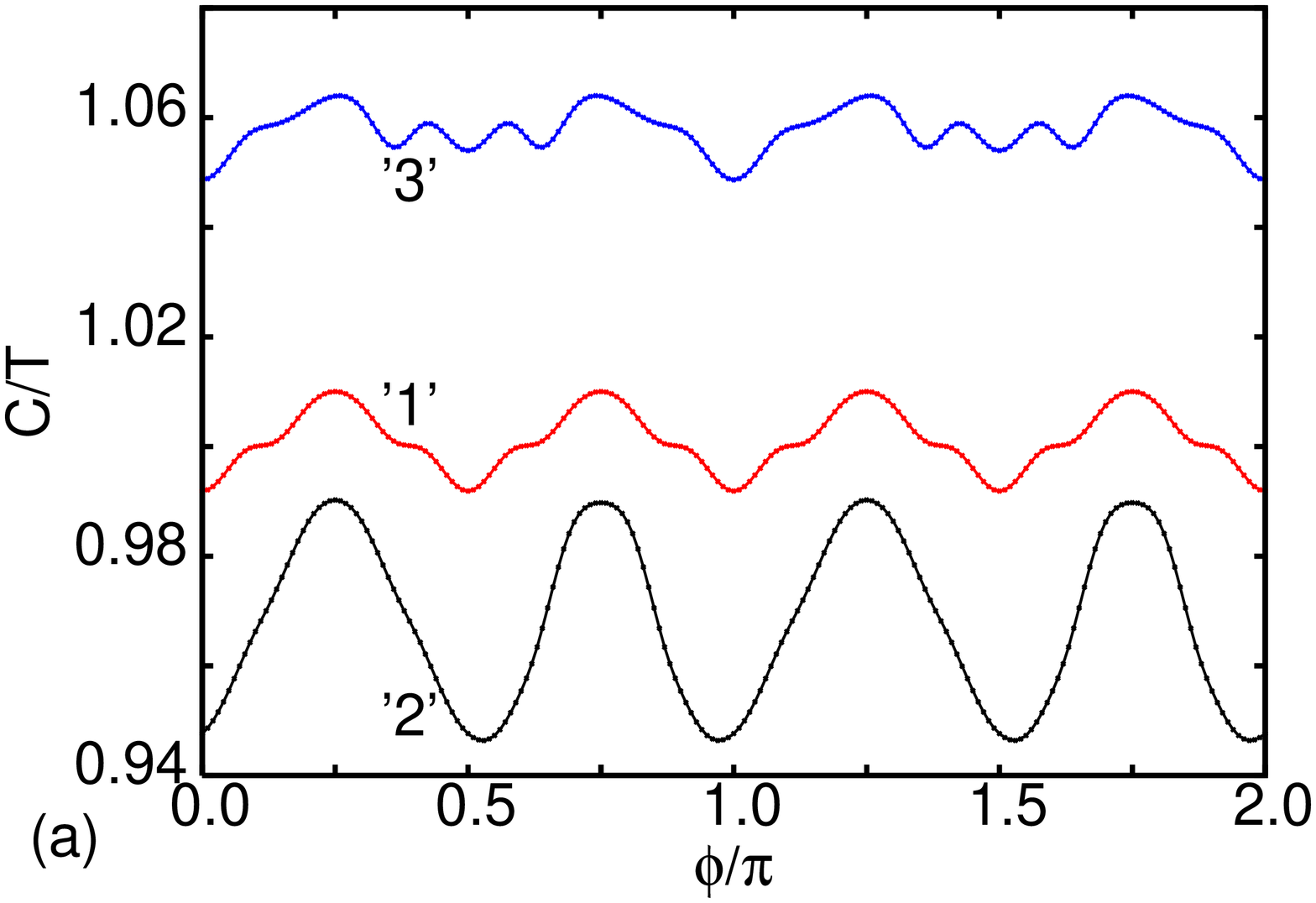,width=6.0cm,angle=-90}
\hspace{-1cm}
\epsfig{file=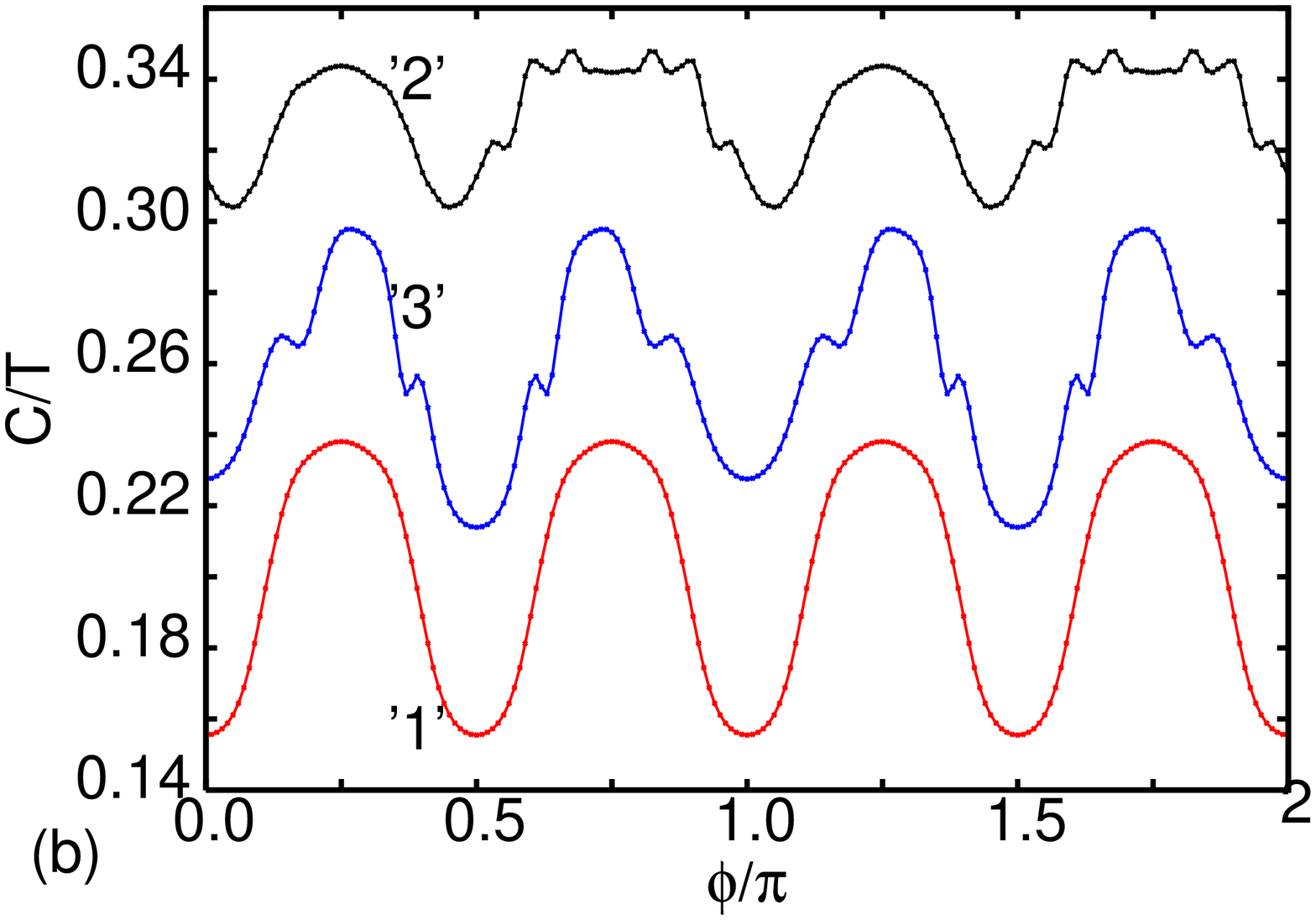,width=6.0cm,angle=-90}
}
\vspace{-0.2cm}
 \caption{
 Specific heat over temperature $\frac{C}{T}$ versus an orientation angle $\phi$ for
the applied Field $H/ \pi \xi H_0=0.005$
at the temperature $T/t=0.006$ (a) and 0.002 (b).
 Note that the different colours  and  numbers '1','2' and '3' correspond
to the three solutions: complex, real and dipole as in Figs. 2 and 3.
 \label{fig4}}
\vspace{-0.2cm}
\end{figure}

The specific-heat results for the three studied states are shown in Fig. \ref{fig3}.
One can see that the lines are close to each other except for a very small temperature 
region where  the complex solution behaves exponentially as in a typical s-wave 
superconducting 
state.

\section{The effect of magnetic field.} 
 
In the presence of a weak in-plane magnetic field the normal electrons
close to the sample surface or vortices experience Doppler shift 
\cite{volovik1985,volovik1993,won2004,won2005,vorontsov2006}.
To take the field into consideration we assume that the penetration depth
$\lambda$ is  fairly larger than the coherence length $\xi$ and shift the 
wave vectors $k_x$ and $k_y$ according to the formulae \cite{tanaka2003}
\begin{eqnarray}
&& \tilde{k}_x = k_x + (H/\pi\xi H_0)\sin\theta,
\nonumber \\
&& \tilde{k}_y=k_y - (H/\pi\xi H_0)\cos\theta,
\label{eq8}
\end{eqnarray}
 where the angle $\theta$ is measured between the $x$-axis and the $B$-field.
The field $H_0=\phi_0/(\pi^2\xi\lambda)$, where $\phi_0=h/2e$ is the 
superconducting flux quantum.  Using the relations (Eq. \ref{eq8}) and previously 
calculated 
$\Delta({\bf k})$ (Fig. \ref{fig1}b) we recalculated the specific heat in the presence of
the magnetic field. 
The corresponding results for different temperatures are presented in Fig. 
\ref{fig4}.
Note that the curves 1 and 2 for the temperature $T /t = 0.006$ (Fig. 
\ref{fig4}a) can be clearly 
related to the 4-fold symmetry, while in the lower temperature $T /t = 0.002$ (Fig.
\ref{fig4}b) 
curves 1 and 3
seem to  follow this symmetry. The other curves 3 for  $T /t = 0.006$  and
2  for $T /t = 0.002$ show a 2-fold symmetry.
Comparing Figs. \ref{fig2} and \ref{fig4} one can conclude that minima in the eigenvalues 
play a similar role as lines of nodes in these curves and the temperature range is very 
important
in such studies. Clearly the specific heat in a lower temperature is more 
sensitive to the type of anisotropy in the eigenvalues and distinguishes easier between 
lines of nodes 
and the simple 
minima in 
the eigenvalue spectra.
For higher temperature the oscillations  $\Delta({\bf k})$ 
 have a fairly small amplitudes, as evident from Fig. \ref{fig1}a. The change of the character of the specific heat 
oscillations is due to the combined effect of (i) smaller  $\Delta({\bf k})$ amplitudes and (ii) smearing of the Fermi 
distribution function as seen from Eq. \ref{eq6}.
Thus the minima and nodal lines are reflected in the angular dependence 
of the specific heat
(caused by the magnetic-field orientation 
with respect to the crystal lattice) in a similar way.

\section{Summary and Discussion}
We have calculated the specific heat of
the 2-dimensional one-band p-wave superconductor 
subjected to 
an external magnetic field of different orientation
 in the lattice plane. This is a commonly used method to identify the 
superconducting gap 
symmetry and nodal states.
The experimental studies of the gap symmetry in different superconductors
by measuring the angle dependence of the specific heat, have been
recently reviewed \cite{park2004}.

Our results show that in different ranges of temperature below 
the superconducting critical temperature $T_c$, the response could be different.
As a matter of fact, the complex solution, showing 4-fold symmetry is reflected in 
the specific-heat
results of the same symmetry, but the two other states can show 2-fold or 4-fold 
symmetry, depending on  temperature. 
In relation to Sr$_2$RuO$_4$, where the 4-fold symmetry has been also found 
\cite{deguchi2004a,deguchi2004b},
we 
conclude that the complex state can be a proper solution for that case.
Note that this kind of solution can be generalized to build a 3-dimensional and 
3-orbital solution of a more realistic model for superconductivity in 
Sr$_2$RuO$_4$, including the  $k_z$-dependence of the superconducting gap $\Delta({\bf k})$
\cite{litak2004,koikegami2003}. That model possesses the horizontal nodal lines
in all three bands which is also consistent with other experiments on the specific heat at 
zero field 
\cite{nishizaki2000} with quadratic dependence of $C$ as a function of temperature.
It has to be remarked that chiral f-wave state have also been proposed to describe some experiments on strontium 
ruthenate \cite{izawa2001,maki2001}.

\begin{acknowledgement}
The  authors acknowledge a partial support by the Polish State
Committee for
Scientific Research (KBN), Project No. 2 P03B 06225 and  
Max Planck Institute for the Physics 
of  Complex Systems in Dresden for hospitality (G.L.).
\end{acknowledgement}

\end{document}